\documentclass[11pt]{amsart}

 \usepackage{tikz}

\usepackage{amsmath}
\usepackage{amsfonts}
\usepackage{amssymb}
\usepackage{amsthm}
\usepackage{hyperref}
\usepackage{cleveref}
\usepackage{graphicx}
\usepackage{enumerate}
\usepackage{cleveref}
\usepackage{mathrsfs}

\usepackage[top=0.9in, bottom=1.15in, left=1.1in, right=1.1in]{geometry}

\hypersetup{colorlinks=true,linkcolor=blue,citecolor=magenta}
\Crefname{conjecture}{Conjecture}{Conjectures}

\theoremstyle{plain}

\theoremstyle{plain}

\title[Mathematics of the MML functional quantizer modules]{Mathematics of the MML functional quantizer modules \\ for VCV Rack software synthesizer}

\author[Schneider--McCarthy--Maxwell--Pfeffer--Schneider--Sills]%{{\bf MTU Mathematics and Music Lab (MML):}\  
{Maxwell Schneider, Cody  McCarthy, Michael G. Maxwell, Joshua Pfeffer,\\ Robert Schneider and Andrew V. Sills}
\address{Department of Computer Science \newline
University of Georgia\newline
Athens, Georgia, 30602, U.S.A.}
\email{maxwell.schneider@uga.edu}

\address{Department of Mathematical Sciences \newline
Michigan Technological University\newline
Houghton, Michigan 49931, U.S.A.}
 \email{cmccarth@mtu.edu}

\address{Department of Visual and Performing Arts \newline
Michigan Technological University\newline
Houghton, Michigan 49931, U.S.A.}
\email{mmaxwell@mtu.edu}

\address{Joshua Pfeffer Graphic Design \newline
Athens, Georgia, 30606, U.S.A.}
\email{josh@joshp.com}

\address{Department of Mathematical Sciences \newline
Michigan Technological University\newline
Houghton, Michigan 49931, U.S.A.}
\email{robertsc@mtu.edu}

\address{Department of Mathematical Sciences \newline
Georgia Southern University\newline
Statesboro, Georgia, 30458, U.S.A.}
\email{asills@georgiasouthern.edu}

\begin{document}
 
 \begin{abstract} 
We detail the mathematical formulation of the   line of ``functional quantizer'' modules developed by the Mathematics and Music Lab (MML) at Michigan Technological University, for the VCV Rack software modular synthesizer platform, which allow synthesizer players to tune oscillators to new musical scales based on mathematical functions. For example, we describe the recently-released MML Logarithmic Quantizer (LOG QNT) module that tunes synthesizer oscillators to the non-Pythagorean musical scale introduced by indie band The Apples in Stereo. 
\end{abstract}

\maketitle

 \thispagestyle{empty}

The {Mathematics and Music Lab} (MML) at Michigan Technological University is an interdisciplinary working group in the departments of  Visual and Performing Arts (VPA) and Mathematical Sciences, sponsored by   professors M. G. Maxwell (VPA) and R. Schneider (Math. Sci.), with student, faculty and industry collaborators. The goals of MML are 
to produce futuristic works of music and installation art, to explore and extend music theory using ideas from mathematics, and to invent new audio hardware and software to bring those projects to life. %MML hosts an array of ambitious theoretical, coding, engineering and music composition projects involving undergraduate and graduate student, faculty, and non-MTU  collaborators. 

In 2007, an unusual musical scale tuned to the logarithms of positive integers was introduced on the album {\it New Magnetic Wonder} by indie band The Apples in Stereo \cite{NMW}. This scale was conceived by the fifth author (R. Schneider), and was subsequently studied in the papers \cite{nonP1, nonP2}, and featured on the 2022 instrumental album {\it Songs for Other Worlds} by Robert Schneider \cite{solo}. It was named the  ``non-Pythagorean'' scale  because the frequencies defining the tones of the scale are irrational numbers bearing little relation to the rational sequence of  frequencies defining the traditional chromatic scale; the discovery of the formula for the chromatic scale is attributed to Pythagoras. {By this definition, many scales used in world musical traditions (e.g.  \cite{world1}) and experimental music \cite{exper1, exper} qualify as  non-Pythagorean; here we focus on a   class of ``functionally quantized''    scales based on mathematical functions with  certain amenable properties.}

MML has developed a line of {\it functional quantizer} modules for the widely-used VCV Rack software   synthesizer platform  \cite{vcv_website}, coded in C++  by MTU Mathematical Sciences master's student C.  McCarthy, based on mathematical formulas derived chiefly by M. Schneider, then a Computer Science and Mathematics double major at University of Georgia. The MML modules allow synthesizer players to play new musical scales tuned to mathematical functions. Our first   functional quantizer module, the MML Logarithmic Quantizer (LOG QNT) \cite{logqnt_site} released in March 2024, modulates control voltage (CV) signals to produce the non-Pythagorean scale in voltage controlled oscillators; see Fig. 1. {\it The MML functional quantizer development team is:}   M. G. Maxwell and R. Schneider (project design), C. McCarthy (lead programmer), M. Schneider and A. V. Sills (mathematical proofs, additional programming), J. Pfeffer (graphics).

Here we describe the mathematical background of the MML functional quantizer line of VCV Rack modules, that make infinite families of mathematical musical scales accessible to modular synthesists. Generalizing the non-Pythagorean scale of the fifth author, a twelve-tone-per-octave scale, let us define a {\it functionally quantized (FQ) musical scale with $T$ tones per octave} as follows. For $0\leq x\leq 1$, let $f(x)$ be a strictly increasing function such that $f(0)=1, \  f(1)=2$, and let $F_0$ denote an arbitrary {\it base frequency} in hertz (Hz) that serves as the root note in the scale. If we identify musical pitch with the corresponding frequency (in hertz), then the $n$th pitch $F_n$ of the first octave of the corresponding FQ scale is defined by the relation
\begin{equation}\label{eq1} F_n=F_0 \cdot f(n/T), \  \  \  \  n=0, 1, 2, 3, \dots, T-1, T.\end{equation}
The  sequence of pitches is repeated in higher/lower octaves by doubling/halving the frequencies.

Twelve tone equal temperament in music theory is the prototype for FQ scales: set $f(x)=2^x,\  T=12,$ such that $F_n=F_0\cdot 2^{n/12}$.  In the logarithmic non-Pythagorean musical scale defined in \cite{nonP1}, the fifth author uses $T=12$ tones playable on a piano keyboard, along with the function
\begin{equation}\label{eq0}
f(x)=\frac{1}{2}\log_2(4+12x),\  \  \  \  x=0,\frac{1}{12}, \frac{2}{12}, \frac{3}{12}, \dots,\frac{n}{12}, \dots, \frac{11}{12}, 1,\end{equation}%\  \  \  \   x=0,\frac{1}{12}, \frac{2}{12}, \frac{3}{12}, \dots,\frac{11}{12}, 1,$$
where $\log_2 t,\  t>0,$ is the base-2 logarithm function, noting $f(x)$ satisfies the boundary conditions $f(0)=1, f(1)=2$ above. Note that we can use  the {\it floor function} $\left\lfloor t\right\rfloor, t\in \mathbb R,$ to rewrite % \eqref{eq0}: % % as % equation: 
\begin{equation}\label{eq0.5} f(x)=\frac{1}{2}\log_2(4+\left\lfloor 12x\right\rfloor),\  \   x\in[0, 1],\end{equation}
since $\left\lfloor Tx\right\rfloor=0,1,2,\dots, T-1, T,$ as $x$ attains the respective values $x=0,\frac{1}{T}, \frac{2}{T}, \frac{3}{T}, \dots,\frac{T-1}{T},$ $1,$ when $x$ moves continuously to the right on the interval $[0, 1]$, which the reader can check. 

Now, in common volt-per-octave CV input calibration for modular synthesizers --- or, more generally, a $V_{\operatorname{ref}}$-volts-per-octave calibration, $V_{\operatorname{ref}}>0$ volts, where $V_{\operatorname{ref}}$ is a {\it reference voltage} ---  let us notate $V_{\operatorname{in}}\geq 0$ volts for the {\it input voltage} and $V_{\operatorname{out}}\geq 0$ volts for the corresponding {\it functionally quantized output voltage} that our LOG QNT module for VCV Rack produces. Set $V_{\operatorname{ref}}=1$ 
volt in usual volt-per-octave calibration; Buchla synths use $V_{\operatorname{ref}}=1.2$ volts. So we have %can write  
 \begin{equation}\label{eq2} F_n=F_0 \cdot 2^{{V_{\operatorname{out}}}/{V_{\operatorname{ref}}}}, \  \  \  \  n=0, 1, 2, 3, \dots, \end{equation}
 where $V_{\operatorname{out}}=V_{\operatorname{out}}(n)$ is the sequence of FQ control voltages yielding the sequence of pitches $F_n$.  
 
 %We note that  $x=\operatorname{frac}(V_{\operatorname{in}}/V_{\operatorname{ref}})$ in the equations above. 
Define the {\it fractional part function} $\operatorname{frac}(t):=t-\left\lfloor t \right\rfloor, \  t\in\mathbb R.$   Then the ``$f(x)$-quantized'' output voltage for a $T$-tone scale is given by the formula 
  \begin{equation}\label{eq3}
V_{\operatorname{out}}\  =\  V_{\operatorname{ref}}\cdot \left\{\left\lfloor {V_{\operatorname{in}}}/{V_{\operatorname{ref}}}\right\rfloor+\log_2f\left(\left\lfloor T\operatorname{frac}\left({V_{\operatorname{in}}}/{V_{\operatorname{ref}}}\right)\right\rfloor\right)\right\}, \end{equation}
as $V_{\operatorname{in}}$ increases, which is obtained by comparing equations \eqref{eq1} and \eqref{eq2}, taking base-2 logarithms and doing a little algebra to solve for  $V_{\operatorname{out}}$. This simplifies in the usual volt-per-octave case  to 
\begin{equation}\label{eq3.5}
V_{\operatorname{out}}\  =\  \left\lfloor V_{\operatorname{in}}\right\rfloor+\log_2f\left(\left\lfloor T\operatorname{frac}(V_{\operatorname{in}})\right\rfloor\right)\   \text{volts}.
 \end{equation}
%which is used for the MML FQNT module for VCV Rack. 
 
For the  logarithmic non-Pythagorean scale  introduced in  \cite{NMW, nonP1}, the FQ output voltage is computed by setting $x=\operatorname{frac}(V_{\operatorname{in}}/{V_{\operatorname{ref}}}),\  T=12$, and $f(x)$ as in  \eqref{eq0.5}  to solve for  $V_{\operatorname{out}}$, yielding %. This yields  the formula   %. This yields the formula    %, % using $f(x)$ as in \eqref{eq0}, with a little algebra,  
%this 
\begin{equation}\label{eq4}
V_{\operatorname{out}}\  =\  V_{\operatorname{ref}}\cdot \left\{\left\lfloor {V_{\operatorname{in}}}/{V_{\operatorname{ref}}}\right\rfloor-1+\log_2\log_2\left(4+\left\lfloor 12\operatorname{frac}\left({V_{\operatorname{in}}}/ {V_{\operatorname{ref}}}\right)\right\rfloor\right)\right\} \end{equation}
% \end{equation}%(We note that if   $V_{\operatorname{in}}/V_{\operatorname{ref}}\geq k> 1$, the $\pm k$ terms shift the voltage down into the 0--1 V range, and the pitches up into the $k$th octave.)
for $V_{\operatorname{in}}\geq 0$ volts. %, after a little algebra. 
In the usual volt-per-octave case,  %$V_{\operatorname{ref}}=1$ volt, 
equation \eqref{eq4} simplifies to 
\begin{equation}\label{eq5}
V_{\operatorname{out}}\  =\  \left\lfloor V_{\operatorname{in}}\right\rfloor-1+\log_2\log_2\left(4+\left\lfloor 12\operatorname{frac}(V_{\operatorname{in}})\right\rfloor\right)\   \text{volts},
 \end{equation}
which is the formula for $V_{\operatorname{out}}$ used in the C++ code for the MML LOG QNT module for VCV Rack \cite{MML_github}. %Equations \eqref{eq3.5} and \eqref{eq5} are compatible with VCV Rack software synthesizer, which uses the volt-per-octave standard. 
Whether $V_{\operatorname{in}}$ is a continuous sweep or an incrementally increasing voltage, formulas \eqref{eq3}, \eqref{eq3.5}, \eqref{eq4} and \eqref{eq5} output a sequence of %functionally quantized 
FQ   voltage values, computed to 6 decimal places of accuracy  in single precision floating point arithmetic (as used in VCV Rack as of this writing).  % in the $k$th octave.
 
Other functional quantizers MML will release for VCV Rack in 2024 (all having $T=12$)  are: % as follows:
\vskip.10in
\begin{enumerate}[(a)]
\item Square Root Quantizer (SQT QNT) with $f(x)=\frac{1}{2}\sqrt{4+12x}, \  0\leq x \leq 1;$
\item Sine Quantizer (SIN QNT) with $f(x)=1+\operatorname{sin}(\pi x / 2), \  0\leq x \leq 1;$
%\item MML Linear Quantizer (LIN QNT) with $f(x)=1+x,   \   0\leq x \leq 1,\  T=12;$  
%\item MML Arctangent Quantizer (ATN QNT) with $f(x)=\operatorname{tan}^{-1} x,\   0\leq x \leq 1,\  T=12;$
\item Power Quantizer (POW QNT) with $f(x)=\frac{1}{2}\left(2^a+\frac{4^a-2^a}{12}x\right)^{{1}/{a}} ,\  0\leq x \leq 1$, $0< a < \infty;$
\item Power Quantizer 2 (POW QNT 2) with $f(x)=1+x^a,\   0\leq x \leq 1$, $ 0\leq a < \infty.$
\end{enumerate}
\vskip.06in
The SQT QNT module was conceived by R. Schneider and A. V. Sills; the SIN QNT module was conceived by M. G. Maxwell and R. Schneider; and  the two-parameter POW QNT modules were conceived by C. McCarthy. %Further MML functional quantizer modules are  in development. %Further details can be found on the MML GitHub server \cite{github}. % (and contains the LIN QNT as the $a=1$ case).%\footnote{We note that the POW QNT module contains LIN QNT as the default case $a=1$, and gives a different square root scale from SQT 
For programming details about the MML modules, see  \cite{MML_github}.
%\newpage

\vskip.15in

\begin{center}\label{fig1}
\includegraphics[scale=.20]{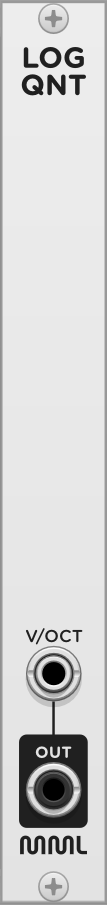} 
\vskip.05in
\small{Figure 1. {\it MML Logarithmic Quantizer (LOG QNT) module for VCV Rack}}
\end{center}
%\vskip.10in

\section*{Acknowledgments}
We are grateful to the Departments of Mathematical Sciences and Visual and Performing Arts at Michigan Technological University for supporting MML and our students; to MTU undergraduates Max Laskaris for helping us name the modules and Josiah Parrott for technical advice about programming for VCV Rack; and to Logan Bayer, Christian Donnell, Martha Sikora and AliAnn Xu,  participants  in the University of Georgia Mathematics and Music Lab sponsored by R. Schneider during 2018--2022, whose undergraduate research on defining, coding and composing with new musical scales  informed  the development of our functional quantizers.

\end{document}